\documentclass[aps,physrev,reprint,superscriptaddress]{revtex4-2}

\usepackage[english]{babel}
\usepackage[utf8]{inputenc}
\usepackage[T1]{fontenc}
\usepackage{graphicx}
\usepackage{amsmath}
\usepackage{amssymb}
\usepackage{orcidlink} 
\usepackage{siunitx}

\usepackage{float}
\usepackage{placeins}

\begin{document}

\title{
 Boundary Element Modeling for THz-SNOM Near-Field Probes
}

\author{Noah Reich\,\orcidlink{0009-0008-7879-3027}}
\affiliation{Ames National Laboratory, U.S.\@ Department of Energy, Iowa State University, Ames, IA 50011, USA.}

\author{Richard Kim\,\orcidlink{0000-0002-4244-9653}}
\author{Thomas Koschny\,\orcidlink{0000-0003-4960-9266}} 
\author{Jigang Wang\,\orcidlink{0000-0002-6159-4119}}
\affiliation{Ames National Laboratory, U.S.\@ Department of Energy, Iowa State University, Ames, IA 50011, USA.}
\affiliation{Department of Physics and Astronomy, Iowa State University, Ames, Iowa 50011, USA.}

\date{\today}

\begin{abstract}
Terahertz scanning near-field optical microscopy (THz-SNOM) enables spatially and spectrally resolved investigation of low-energy collective and topological states in quantum materials and nano-structured devices by exploiting the strongly localized near field generated by a polarized atomic force microscopy (AFM) tip.
Because the THz near-field response is highly sensitive to tip geometry, understanding how geometric parameters control field confinement and decay is essential for selecting tip designs that provide the spatial resolution required for specific experiments.
In this work, we develop a boundary element method (BEM) for modeling the electrostatic near-field of a conducting AFM tip in response to an external background field. 
The tip is represented by a simplified closed, blunt-cone geometry with its surface discretized into panels using increased mesh density near the apex. 
A modified Green’s function is used to calculate interactions between surface panels, while an augmented linear system simultaneously enforces the equipotential boundary condition on the conducting surface and overall charge neutrality to determine the induced surface-charge distribution.
We validate the BEM by comparing the resulting near-field decay with that obtained from a three-dimensional quasi-static COMSOL Multiphysics simulation. 
The BEM accurately reproduces the COMSOL field-decay behavior while requiring orders of magnitude fewer mesh elements and substantially reducing computational cost. 
These results demonstrate that the BEM provides an accurate and computationally efficient framework for systematically investigating the dependence of THz near-field confinement on AFM tip geometry. 
This framework can be extended to more complex and experimentally realistic tip geometries, providing a route toward rational tip selection and design for application-specific spatial-resolution requirements.
\end{abstract}

\maketitle

\section{Introduction}
Scattering-type Scanning Near-field Optical Microscopy (sSNOM) is an essential experimental technique that uses the nanometer spatial resolution of an Atomic Force Microscope (AFM) illuminated by visible to infrared light of about half a micron to ten's of micrometers wavelength to gather chemical and optical properties from material samples at length scales far below the diffraction limit associated with the wavelength of the illuminating probing light \cite{knoll,hill,qazi,chen,fei,nrm,zrte5}.
At much longer wavelengths, Terahertz frequencies are able to probe low-energy electronic behavior that is often inaccessible to visible or mid-infrared light while revealing physical information about low-energy collective states and topological edge states in quantum materials or resonant current distributions in deeply sub-wavelength metasurfaces or nano-structured devices \cite{apr,htchen,ribb,zhang,stin,agha,pizz,n4,n5,kim2:2023,kim1:2023,n6,apl,mlg, Haeuser2026Terahertz,Haeuser2024Analysis}.
In particular, the recent development of a sub-2 K cryogenic magneto-THz scanning near-field microscopy platform overcomes the conventional liquid-helium temperature limit, enabling spatially resolved measurements of low-energy electrodynamic excitations and collective responses in quantum materials deep within their low-temperature quantum phases and under applied magnetic fields \cite{Kim2023Cryogenic,Haeuser2024Analysis,Haeuser2026Terahertz,Kim2026Graphene}.

\begin{figure}[t]
    \centering
    \includegraphics[width=0.99\linewidth]{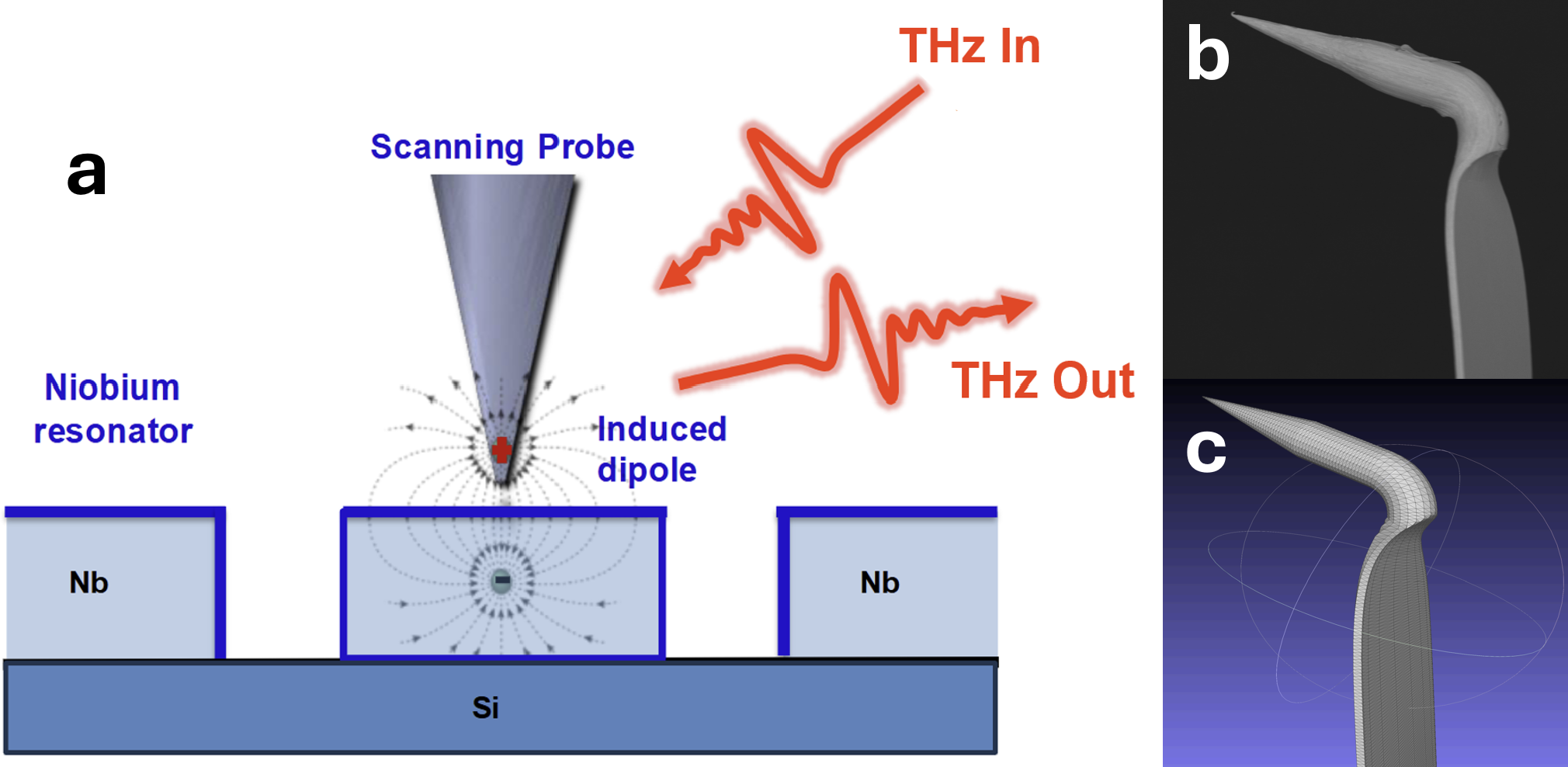}
    \caption{%
    Schematic of a typical SNOM experiment and cantilever-type AFM tip geometry.
    (a) Illustration of an experimental THz-SNOM measurement displaying the induced dipole between the scanning polarized AFM tip and a structures surface of a sample below.
    (b) Scanning Electron Microscopy (SEM) picture of the sharp conical section of a commercial AFM tip/cantilever (Rocky Mountain Nanotechnology 25Pt300B) used for THz sSNOM.
    (c) 3D STL mesh model recreation of SEM probe picture that can be imported into simulation domains.
    }    
    \label{fig:SNOM-schematic}
\end{figure}

As illustrated in Figure \ref{fig:SNOM-schematic}(a), an incident THz pulse polarizes the AFM tip, which drives charges along the surface and concentrates them near the bottom apex. 
This creates a large field enhancement below the tip that also induces charges on a sample beneath the apex. 
These effects come together to create an induced electric near-field decay on the surface of the sample. 
This is an oscillating dipole-like system between the now-coupled tip and surface, which radiates a scattered THz signal that is experimentally measured (Figure \ref{fig:SNOM-schematic}(a)). 

The geometry of the AFM tips directly affects the near-field decay between the AFM tip and the sample, since the differing tip shapes have different near-field responses to an initial excitation. 
The geometry can affect spatial resolution, field confinement, field penetration, and contrast strength \cite{Pistore2024THzProbes,Maissen2019THzProbes,Mastel2017THzTips,Jiang2026THzProbeOptimization}. 
Therefore, the response of different tip geometries needs to be studied so that an ideal shape can be selected for varying experimental needs. 
Here, we construct and validate a boundary element method (BEM) simulation currently designed for axisymmetric, closed tip boundary shapes. 
Tip parameters are fixed for the study, as investigating the near-field geometrical dependencies is beyond the scope of this report. 
Instead, the focus is to create the foundations of a method that will inevitably be used to investigate these dependencies. 
This work details the methodology of the simulation approach, all aspects of the setup of the BEM simulation, results comparing the method against another common simulation technique, and a conclusion that highlights the findings and how the constructed simulation will be taken further.

\section{Methodology}

\subsection{Problem Statement}
Because in a typical THz-sSNOM the experimental AFM tip ($\sim 80\mathrm{\mu m}$ in our case) and, in particular, the extent of the probing near field between the nanoscale apex of the tip and the sample surface (below $\sim 100\mathrm{nm}$) are substantially smaller than the wavelength of the illuminating THz beam ($\sim 300\mathrm{\mu m}$ at 1 THz), interaction of tip and sample is essentially an electrostatic (quasielectrostatic) problem.
This holds especially for the many order of magnitude subwavelength space between the sharp apex of the AFM tip and the sample surface, which is responsible for the readout signal generated by the instrument. 
Here all retardation effects from free-space propagation can consequently neglected and the field distribution becomes purely an electrostatic problem.  
This holds despite the fact that the overall tip/cantilever assembly (as well as our analytical tip-shape modes are only marginally subwavelength and, in principle, would show some retardation effects and corresponding resonances in the global response at the large scale.

The actual near-field response to an incident THz pulse between tip and sample can be simulated with the following problem statement: 
A neutral conductor is placed in an external, homogeneous background field,
\begin{equation}
    \vec{E}_{background} = E_0\,\hat{z}, \quad
    \phi_{background} = -E_0\,z.
    \label{eq:background}
\end{equation}
Charges on the conducting surface will align themselves so that there is no tangential electric field on the surface or, equivalently, the surface is equipotential. 
This creates a surface charge density that heavily depends on the geometry of the tip and is static, as a zero surface tangential electric field denotes the absence of force on the surface charges. 
The induced surface charge creates an induced potential and electric field as a byproduct. 
The near-field decay of this induced electric field can then be evaluated to find geometrical dependencies of its decay shape. 
The resulting electrostatic potential is governed by Poisson's equation,
\begin{equation}
    \nabla^2\phi = -\varrho(\boldsymbol{r})/\varepsilon_0
    \label{eq:poisson_intro}
\end{equation}
which provides the starting point for the boundary-integral derivation developed below.

\subsection{Boundary Element Methods}
Simulating tip responses can be done through many simulation techniques. 
A standard method is using Finite Element Methods (FEM) to solve for the tip's response and physical properties in a surrounding domain using a full 3D simulation software such as COMSOL Multiphysics \cite{COMSOL_ES_64}. 
While convenient, this simulation method comes with many downsides, such as requiring high computational power, large memory, and an expensive license. 

One way to address these issues is to use Boundary Element Methods (BEM) \cite{McLeod2014BEM,Shen2008BEM} instead. 
This technique meshes only the surface of a simulated object by dividing the boundary into small sections called 
panels allowing one to avoid meshing the surrounding free space
\cite{BrebbiaDominguez1992,BeerSmithDuenser2008,ChengCheng2005,Steele1997}.
The physics of the simulation is then solved only at the panel midpoints. 
The result is a method that is efficient, computationally inexpensive, and can be run for free using a few lines of Python or simple C code. 

\subsection{Green's Function}
There are still many possible approaches to formulating a BEM technique \cite{ChengCheng2005,SauterSchwab2011}.
The method selected here is using Green's function \cite{Kellogg1929}, which accounts for panel-to-panel interaction or how an individual surface charge affects all other surface charges on the conducting surface.
This function is analytically known for a reasonably smooth surface, independent on the geometric shape.
This is relevant to the physical setup as each surface charge will affect the locations of all other surface charges until a static-equilibrium state is found, which is the condition required by conductor properties. 
The free-space Green's function for the 3D Laplacian is given by
\begin{equation}
 G(\boldsymbol{r};\boldsymbol{r'}) = -\frac{1}{4\pi\,|\boldsymbol{r}-\boldsymbol{r'}|},
 \label{eq:greensfunction}
\end{equation}
which satisfies 
$\nabla^2G(\boldsymbol{r};\boldsymbol{r'})=\delta(\boldsymbol{r}-\boldsymbol{r'})$.

\section{Setup}
\subsection{Surface Parameterization}
As stated in a previous section, the geometry of the AFM tip, especially around the sharp apex, significantly changes the resulting induced charge density distribution on AFM tip and sample surface when polarized by the external THz field and influences the spatial decay of the near-field away from the tip, which determines the behavior of the approach curves. 
The shape selected for this study is a simplified geometry derived from SEM pictures of physical AFM tips as used in experiments and is shown in Figure \ref{fig:SNOM-schematic}(b,c).
Here, we concentrate on the shape of the tip alone, the sample surface is not included as an electromagnetic boundary. 
We isolate the field generated by polarization of the tip itself and characterize spatial decay of the near-field below the apex. 
Tip–sample coupling, image-charge effects, and material-dependent sample response are therefore outside the present model.
The surface of our tip model includes three smoothly transitioning parametric curves: 
a curvature radius $(R_c)$, a conical body, and a spherical cap $(R_0)$, illustrated in Figure \ref{fig:parametric-tip-geometry}. 
Benefits of this simplified design include axial symmetry and a closed body, both of which considerably simplify the methodology. 
The axial symmetry effectively turns the surface into a rotation shape given by a surface curve.

The surface curves are parameterized in the form of $\rho$ as a function of $z$ to have functions that describe the radius of the tip based on the vertical location.
Here, the $z_n$ locate the transition points in the surface curve, $z_0$ is the apex of the model tip with the nanoscopic curvature radius $R_c$, at $z_1$ the spherical tip transitions into a tangential cone, which, at $z_2$ transitions tangentially into the large spherical end cap with curvature radius $R_0$.
In the experimental tip, this would transition smoothly into the cylindrical wire with radius $R_0 \approx 30\mathrm{\mu m}$ that then higher up bends and flattens into the flat beam-shaped cantilever as shown in Figure \ref{fig:SNOM-schematic}(b,c), but for our purposes here, we only consider the more compact and axisymmetric model (Figure \ref{fig:parametric-tip-geometry}). 
We get piecewise parametrization
\begin{equation}
 \begin{aligned}
 z'(u) &= u, \ \forall u \in [z_0,z_3], \\[4pt]
 \rho'(u) &=
  \begin{cases}
   \sqrt{R_c^2-\left[R_c-(u-z_0)\right]^2},  & z_0 \leq u \leq z_1, \\[4pt]
   R_c \cos\alpha \,+\, (u-z_1)\,\tan\alpha,          & z_1 \leq u \leq z_2, \\[4pt]
   \sqrt{R_0^2-\left[u-(z_3-R_0)\right]^2},  & z_2 \leq u \leq z_3,
  \end{cases}
 \end{aligned}
  \label{eq:parametric_curves}
\end{equation}
for the sharp tip section with curvature radius $R_c$, the conical shaft, and the larger spherical end cap with radius $R_0$, respectively.
The vertical segment are given in terms of $R_c$, $R_0$, and the cone half-angle $\alpha$ by
\begin{equation}
 \begin{aligned}
        z_0 &= 0, \\
        z_1 &= z_0+R_c\left(1-\sin\alpha\right), \\
        z_2 &= z_1+\left(R_0-R_c\right) (\cos\alpha)\,\cot\alpha, \\
        z_3 &= z_2+R_0\left(1+\sin\alpha\right),
 \end{aligned}
\end{equation}
determining the length of the tip structure.
\begin{figure}
    \centering
    \includegraphics[width=0.99\linewidth]{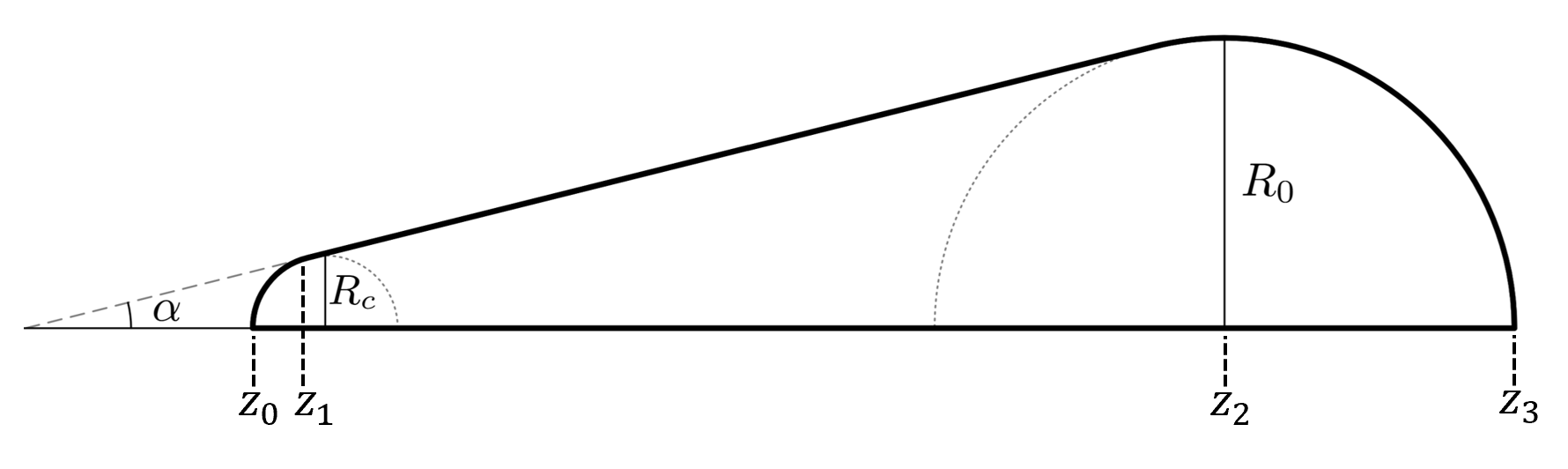}
    \label{fig:parametric-tip-geometry}
    \caption{Outline of the simplified axial-symmetric tip geometry model based on SEM images in the $(z,\rho)$-plane.}
    \label{fig:parametric-tip-geometry}
\end{figure}

\subsection{Governing Equations}
Two equations govern the physics of the setup. 
The first is the surface boundary condition of the conductor.
\begin{equation}
    \phi_{ind}+\phi_{bkg} = \phi_{total} = V_{tip}
    \label{eq:surfaceboundarycondition}
\end{equation}
This equation results from equipotential conducting surfaces. 
Having a constant, homogeneous background electric field translates to a linear background potential. 
The induced potential on the surface must then cancel out all spatial variation of the background potential and leave an equipotential surface with constant potential $V_{tip}$.

It was initially hypothesized that the problem's physical setup causes the charge density to be symmetric by nature, which would force $V_{tip}$ to automatically equal zero for all conductor shapes. 
This creates an issue, as shifting the conductor around in the linear background potential changes the strength of the induced potential. 
An equivalent action is scaling the surface boundary condition, as the infinity condition, $\phi_{ind}\rightarrow 0$ as $r\rightarrow\infty$, remains unchanged. 
This creates a net charge on the conductor surface, which adds a monopole contribution to the existing dipole contribution of the induced surface charge density. 
Electric dipoles and monopoles have different field decay; therefore, spatially shifting the conductor in the background field, while forcing $V_{tip}$ to be zero, changes the resulting induced field decay. 
This thought process concludes that there is one value for $V_{tip}$ that results in a neutral conductor. 

The second governing equation comes from the problem statement of placing a neutral conductor in a background field; therefore, the net charge on the conducting surface must be zero after initial polarization,
\begin{equation}
    Q_{total} = \int_S\sigma \,dS = 0
    \label{eq:totalcharge}
\end{equation}
This idea pairs with the first condition as there is a value of $V_{tip}$ that will result in a balanced charge distribution, the value of which depends on the geometry of the simulated conductor. 
By introducing the neutrality condition, the system will solve for $V_{tip}$ and insert it into the boundary condition of the surface. 

\subsection{Analytical Setup}
The nature of electrostatics creates the condition that a potential in a charge-free domain must satisfy Laplace's equation, which we can enforce on the target induced potential outside the polarized tip. Therefore, we can start with an applied Poisson's equation
\begin{equation}
    \nabla^2\phi(\boldsymbol{r};\boldsymbol{r'}) = -\frac{\varrho}{\epsilon_0}=-\frac{\sigma}{\epsilon_0}\delta(\boldsymbol{r}-\boldsymbol{r'})
    \label{eq:poisson2}
\end{equation}
Then, we plug in and use the linearity of the relevant Green's function (\ref{eq:greensfunction}) to get
\begin{align}
    \phi_{ind}(\boldsymbol{r}) &= 
    \int G(\boldsymbol{r},\boldsymbol{r'}) \,\sigma(\boldsymbol{r'}) \,d\boldsymbol{r'} 
    \nonumber \\
    &=
    \frac{1}{4\pi\epsilon_0}\int_S\frac{\sigma(\boldsymbol{r'})}{|\boldsymbol{r}-\boldsymbol{r'}|} \,dS'
    \label{eq:phiind1}
\end{align}
where $dS$' is a 3D surface element at the surface parameter $(u',\phi')$
\begin{equation}
    dS' = \rho'(u')\, \sqrt{\left(\frac{dz}{du'}\right)^2+\left(\frac{d\rho}{du'}\right)^2} du'\, d\phi'
    \label{eq:surfaceelement}
\end{equation}
The induced potential equation (\ref{eq:phiind1}) needs to be converted into cylindrical coordinates
\begin{equation}
    \begin{gathered}
    \phi_{ind}(\rho,z,\phi) \ =\ 
    \frac{1}{4\pi\epsilon_0} \iint \,dS'(u',\phi') \hspace{20mm} \\ 
    \frac{\sigma(u')}{\sqrt{\rho^2+\rho'^2-2\rho\rho'cos(\phi-\phi')+(z-z')^2}}  
    \end{gathered}
    \label{eq:phiind2}
\end{equation}
Since the surface is axisymmetric, only the relative azimuthal dependencies affect the solution; therefore, we can set $\phi-\phi'$ equal to a single variable and integrate out this variable in the above \cite{Provatidis1998}. 
This process results in
\begin{equation}
    \begin{gathered}
        \phi_{\textit{ind}}(\rho,z) \ =\ 
        \frac{1}{\pi\epsilon_0} \int \frac{\sigma(u')\,\rho'(u')}{(\rho+\rho')^2+(z-z')^2}\,
        \\K\left(\sqrt{\frac{4\rho\rho'}{(\rho+\rho')^2+(z-z')^2}}\right) \\
        \cdot \,\,\sqrt{\left(\frac{dz'}{du}\right)^2+\left(\frac{d\rho'}{du'}\right)^2}\, du'
    \end{gathered}
    \label{eq:phiind3}
\end{equation}
where $K(k)$ is the complete elliptic integral of the first kind. 
This special function makes reducing equation (\ref{eq:phiind3}) further non-trivial; 
however, there are numerical strategies that can be applied to solve the equation \cite{Kress2014}. 

From this point on in the derivation, the resulting equation becomes a one-dimensional integral along the surface curve of the conductor and has only one unknown: the surface charge density. 
Plugging in the induced potential on the surface from the boundary condition (\ref{eq:surfaceboundarycondition}) allows us to solve for the unknown charge distribution.
\begin{equation}
    \begin{gathered}
          \phi_{ind}\big(\rho(u),z(u)\big) 
          \Big|_{\mathrm{surface}} 
           =\ E_0\,z(u)+V_{tip}
    \end{gathered}
    \label{eq:phiind4}
\end{equation}
Here, \(u\) parametrizes the observation point along the generating curve of the conductor surface, whereas \(u'\) in (\ref{eq:phiind3}) parametrizes the source points over which the surface integral is evaluated. 
Writing the observation coordinates as \((\rho(u),z(u))\) restricts the field point to the conductor surface, allowing the equipotential boundary condition to be enforced point-wise along the boundary. 
Lastly, the surface parametric equations (\ref{eq:parametric_curves}) can be inserted into (\ref{eq:phiind4}) to convert equation (\ref{eq:phiind3}) into a solvable form. 

\subsection{Surface Discretization}
A key characteristic of BEM is to discretize only the surface with panels (mesh points) in a way that suits the simulation objective instead of the volume surrounding it. 
Radial and axial electric field decay curves under the apex are the target outputs of the simulation as material samples sit beneath the tip. 
Because of this, the surface mesh is built with more panels along the bottom of the tip. 
The first step of this process is to split N panels between the three surface curves using fractional distribution. 
The current setup allocates 47$\%$ to the curvature radius, 45$\%$ to the conical body, and 8$\%$ to the spherical cap. Using a power law grading on the panel lengths per segment can further concentrate the mesh, as a higher power grading will reduce the panel size toward the bottom of each segment by defining the panel edges to be
\begin{equation}
    \begin{aligned}
        z_i &= s^{-1}\left[\left(\frac{i}{N_\mathit{segments}}\right)^p S \,\right], \\
        s(z) &= \int_{z_\mathit{left}}^z 
                 \sqrt{\left(\frac{dz}{du}\right)^2 + \left(\frac{d\rho}{du}\right)^2} du,
    \end{aligned}
    \label{eq:sfunction}
\end{equation}
where $z_i$ for $i=0,1,...,N_{segments}$ is the vertical location of $i$-th edge, 
$S$ is the total arc length of the segment, and $p$ is the grading power.
$s(z)$ is the cumulative arc length measured along the surface curve from the beginning of the segment, $z_{\mathrm{left}}$, to the axial position $z$. 
The inverse function $s^{-1}$ maps a specified arc-length position back to the corresponding axial coordinate $z$ along the surface.

The grading powers for the curvature radius, conical body, and spherical cap used are 2, 1.7, and 1, respectively. A grading power of 1 simply gives linear spacing throughout the segment, and a higher grading power will result in smaller panels and a more concentrated mesh toward the bottom of the segment. 

\begin{figure}[t]
    \centering
    \includegraphics[width=0.99\linewidth]{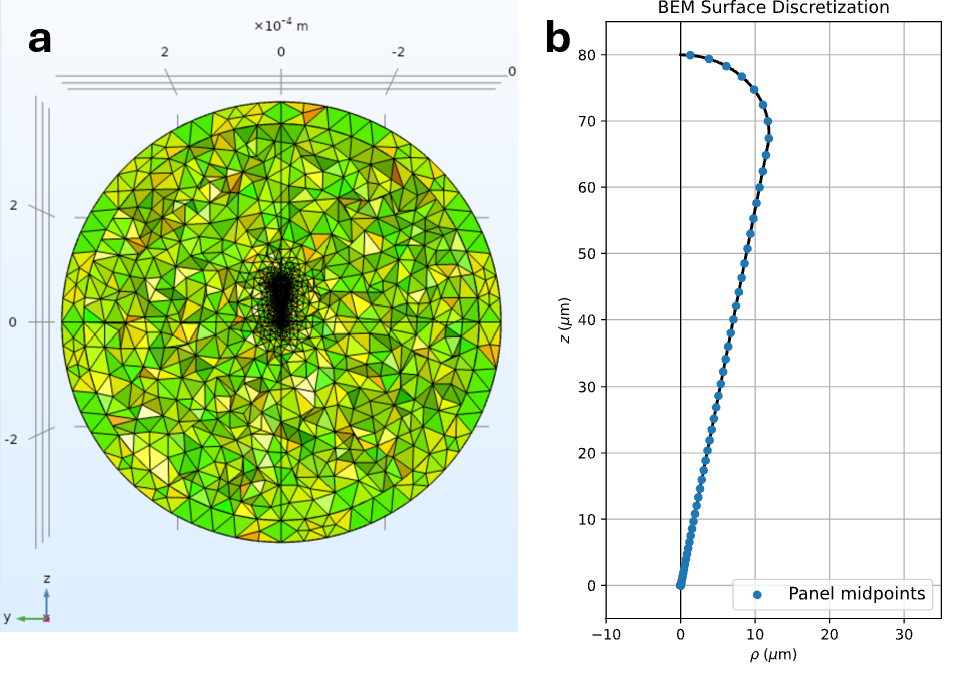}
    \caption{%
    Mesh comparison between the 3D COMSOL simulation and the discretization of the BEM surface model. 
    (a) Volume mesh of quasi-static COMSOL simulation that is set to closely match the BEM definition. 
    (b) Surface discretization of the BEM method of 100 panels fractionally distributed to the three surface curves.
    }
    \label{fig:mesh-comparison}
\end{figure}

\subsection{Numerical Setup}
To numerically solve for the unknown surface charge density, the boundary condition (Eq. \ref{eq:surfaceboundarycondition}) is forced on all panel midpoints, called collocation points, of the discretized surface with a key assumption that the surface charge is constant within each panel. 
A matrix equation can then be constructed to solve for the $j$-th panel surface charge and global surface potential.

The $i$-th collocation point is defined by the surface parameter $u_i$, with coordinates
\begin{equation}
    (\rho_i,z_i) = \big( \rho(u_i), z(u_i) \big),
    \label{eq:discretizationpoints}
\end{equation}
and the $j$-th source panel span the surface-parameter interval $u' \in [u_j^{-},u_j^{+}]$.

The influence coefficient $A_{ij}$ is then defined as
\begin{equation}
\begin{aligned}
A_{ij}&=\frac{1}{\pi\epsilon_0}\int_{u_j^-}^{u_j^+}\frac{\rho(u')}{\sqrt{\left[\rho_i+\rho(u')\right]^2+\left[z_i-z(u')\right]^2}}
\\&\qquad \cdot K\left(\sqrt{\frac{4\rho_i\,\rho(u')}{\left[\rho_i+\rho(u')\right]^2+\left[z_i-z(u')\right]^2}}\right)
\\&\qquad\cdot\sqrt{\left(\frac{dz(u')}{du'}\right)^2+\left(\frac{d\rho(u')}{du'}\right)^2}\,du',
\end{aligned}
\label{eq:Acomponents}
\end{equation}
where $A_{ij}$ describes the induced potential at collocation point $i$ due to a unit surface charge density on source panel $j$. 
The coefficients $A_{ij}$ form the $N\times N$ influence matrix
\begin{equation}
    \mathbf{A} =\left[A_{ij}\right]\in \mathbb{R}^{N\times N};
    \quad i,j=1,\ldots,N.
    \label{eq:Amatrix}
\end{equation}

The surface area of the $j$-th panel is given by
\begin{equation}
 S_j = 2\pi \int_{u_j^-}^{u_j^+} \rho(u')
 \sqrt{\left(\frac{dz(u')}{du'}\right)^2+\left(\frac{d\rho(u')}{du'}\right)^2}\,du',
 \label{eq:Sj}
\end{equation}
and the panel surface areas are collected into the vector
\begin{equation}
    \mathbf{S}^{T}= \begin{bmatrix}
         S_1 &S_2 &\cdots &S_N
    \end{bmatrix}.
    \label{eq:ST}
\end{equation}

The discretized surface charge density and collocation-point axial coordinates are similarly written as
\begin{equation}
 \begin{aligned}    
  \boldsymbol{\sigma} &=
   \begin{bmatrix}
    \,\sigma_1 & \sigma_2 & \cdots & \sigma_N \,
   \end{bmatrix}^T \\
  \mathbf{z} &=
   \begin{bmatrix}
    \,z_1 & z_2 & \cdots & z_N \,
   \end{bmatrix}^T 
 \end{aligned}
 \label{eq:sigma_z}
\end{equation}
The resulting augmented matrix equation is
\begin{equation}
\begin{bmatrix}
    \mathbf{A} & -\mathbf{1} \\
    \mathbf{S}^{T} & 0
\end{bmatrix}
\begin{bmatrix}
    \boldsymbol{\sigma} \\
    V_{\mathrm{tip}}
\end{bmatrix}
\ =\
\begin{bmatrix}
    E_0\,\mathbf{z} \\
    0
\end{bmatrix}
\label{eq:matrixequation}
\end{equation}
where $\mathbf{1}$ is an $N\times1$ vector of ones. 
Equation (\ref{eq:matrixequation}) is a matrix equation that numerically decomposes, row-wise, into the following two equations for the induced potential and the overall charge of the tip, which need to be solved simultaneously to find the unknown $\sigma_j$,
\begin{align}
\phi_{\mathrm{ind}}(\rho_i,z_i) 
 &\ =\ \sum_{j=1}^{N} A_{ij}\sigma_j - V_{\mathrm{tip}}
 \ \stackrel{!}{=}\ E_0\, z_i, 
 \label{eq:phiind_Q_1}\\
Q_{\mathrm{tot}} 
 &\ =\ \sum_{j=1}^{N}\sigma_j S_j
 \ \stackrel{!}{=}\ 0,
\label{eq:phiind_Q_2}
\end{align}
where the surface charge density $\boldsymbol{\sigma}$, $\mathbf{z}$, and $-\mathbf{1}$ are all $N\times1$ matrices, $\mathbf{S}^{T}$ is a $1\times N$ matrix, and $V_{\mathrm{tip}}$ is a global scalar.

A barrier when solving for the charge density of each panel comes from a logarithmic singularity of the first elliptic integral when its input trends towards one. 
For the current formulation, the singularity transpires when the observation and source point are identical.
Solving the matrix $(\mathbf{A})$ accounts for how a panel affects its own location, which results in the singularity. 
To avoid this numerical error, the integral $(A)$ is split around the logarithmic singularity if the observation point lies inside the current source panel. 
This is done by calculating the integral from the bottom of the panel to a small shift below the panel midpoint, and then a small shift above the midpoint to the top of the panel. 
These two shifted calculations are then summed together and taken as the integral value for the source panel. 

The induced potential contribution for each panel can then be calculated by plugging the panel surface charge density into Equation (\ref{eq:phiind3}). 
An important addition to the formulation is that, during the potential calculation, each panel is split up into smaller Gauss-Legendre quadrature points. 
Although the surface charge density is assumed to be constant on each panel, the induced-potential integrand varies across the panel because the Green’s function and geometric terms depend on the source location. 
The Gauss–Legendre points therefore allow this spatial variation to be integrated accurately without introducing additional surface charge unknowns or requiring further refinement of the BEM mesh. 

Finite-difference approximation is then used to calculate the induced electric-field components from potential values evaluated at nearby shifted observation points. 
Each potential evaluation uses the Gauss–Legendre quadrature points to integrate the contributions from all surface charge panels. 
Although the electric field could be calculated directly by differentiating the Green’s function integral, finite differences provide a simpler method that avoids directly deriving and integrating the spatial derivatives of the kernel. 
The final output of the simulation is induced field decay plots that evaluate the electric field contributions at 320 points along the radial and axial lines. 
It is an important distinction to make that although the output depends on the number of quadrature and cut line evaluation points, neither addition adds more mesh points to the setup, as they don't add more unknowns, do not enforce the boundary condition, and do not interact with each other. 
They are simply used to approximate an integral and as observation points.
Numerical calculations were performed using the SciPy library \cite{Virtanen2020SciPy}, including routines to solve linear systems arising from the matrix equation (solve), one-dimensional definite integration (quad), and complete elliptic integrals (ellipk).

\subsection{Comparison with Quasi-static FEM simulation}
The main motive of constructing a BEM simulation is to prove that, for the local near-filed between tip and sample surface, it is both as reliable, and it is more efficient and accessible than a traditional FEM simulation, which can be done by directly comparing the BEM results to a matching COMSOL simulation. 
This comparison can be taken further by comparing to a quasi-static COMSOL simulation \cite{COMSOL_EMW_64} with an analogous physical setup, which can both solidify the benefits of a BEM simulation and confirm the setup of the BEM electrostatic physics.
Note that this also shows that despite possible differences due to retardation effects on the global scale of the AFM tip ($\sim 80\mathrm{\mu m}$ $\lesssim$ $\lambda\approx 300\mathrm{\mu m})$ the actually relavant near-field between tip and sample is faithfully reproduces in both electrostatic BEM and plane wave scattering in quasi-electrostatic FEM in COMSOL. 
The COMSOL setup, as can be inferred from the FEM mesh shown in Figure~3(a), includes the 3D tip geometry whose surface is set to a perfect electric conductor (PEC) in the center and surrounded by a spherical domain set to the wave equation of air. 
A relatively small perfectly matched layer (PML) is added to the outside of the domain \cite{COMSOL_PML_64}, with the outer spherical surface terminating the computation domain also being set to a PEC. 
A linear THz plane wave with the E-field polarized in the z-direction and the wave vector perpendicularly is then sent at the tip as a background field, which is analogous to placing the conductor in a homogeneous background field. 

To be comparable, the COMSOL mesh across the surface of the tip now needs to match the BEM discretization. 
The fractional distribution along the three curve segments constituting the cross-section of the axisymmetric tip model is accounted for by designating the mesh size of a curve to
\begin{equation}
    \text{mesh size} \ =\ \frac{\text{segment arc length}}{N_\mathit{segments}}
    \label{eq:meshsize}
\end{equation}
The result is a 3D meshed tip whose surface mesh points closely reflect that of a revolved BEM tip. 
The domain mesh is set so that the axial and radial cut lines have roughly the same number of data points/mesh points as the BEM cut lines.

\begin{figure}[t]
    \centering
    \includegraphics[width=0.99\linewidth]{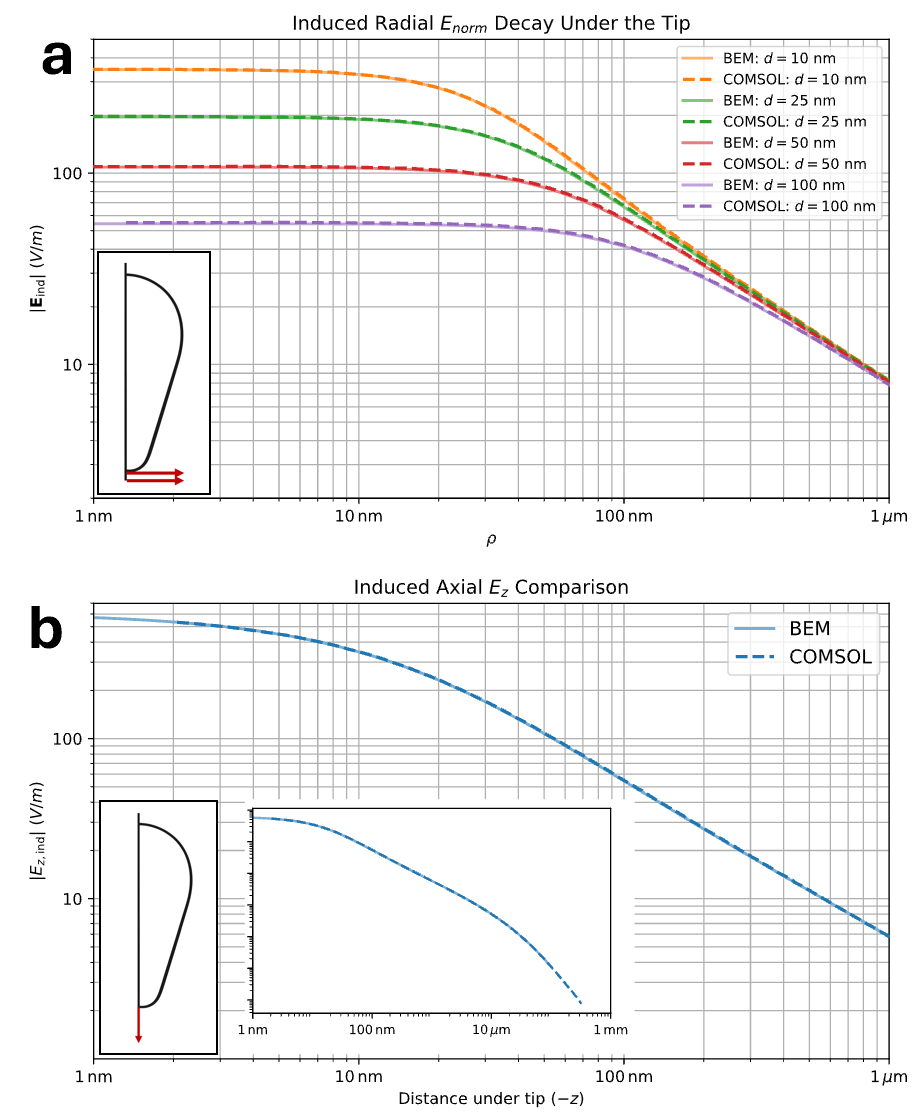}
    \caption{%
    Radial and Axial induced field decay under the tip apex ($R_c=30\text{nm}$) comparison.
    (a) Induced radial decay at varying distances beneath the apex.
    (b) Induced axial $|E_z|$ decay beneath the apex. The inset plot displays the full data line to show the three decay regions: constant plateau, intermediate region, and dipole behavior $\sim1/r^3$.}
    \label{fig:decay-plot-comparison}
\end{figure}

\section{Results}
The number of mesh elements, 1D curve elements for the BEM and 2D surface and 3D volume elements for the FEM, are quantified in Table 1, which reveals the magnitudes of the increased mesh used in the 3D simulation. 
The resulting volume mesh of the COMSOL simulation, shown in Figure \ref{fig:mesh-comparison}(a), is incredibly dense near the tip, as the domain mesh requires smoothing out the refined tip mesh into the more sparse domain mesh. 
Magnitudes of unnecessary mesh points are a byproduct of the 3D mesh, as the valued output is only the field decay along 1D lines beneath the apex. 
Contrarily, the surface discretization of the BEM method shown in Figure \ref{fig:mesh-comparison}(b) produces a mesh that has zero unwanted domain mesh points and whose concentrated mesh points minimally reflect the desired output of the field beneath the tip. 
Table 1 also displays that the BEM method is around 28 times faster than traditional FEM.

\begin{table*}[t]
\caption{Comparison of BEM and COMSOL mesh requirements and computation times.}
\label{tab:mesh_comparison}
\centering
\renewcommand{\arraystretch}{1.3}

\begin{tabular*}{0.9\textwidth}{@{\extracolsep{\fill}} l c c c c @{}}
\hline\hline
Method &
Tip elements &
\begin{tabular}{@{}c@{}}Domain elements\end{tabular} &
\begin{tabular}{@{}c@{}}Total mesh elements\end{tabular} &
\begin{tabular}{@{}c@{}}Computation time \\ (seconds)\end{tabular} \\
\hline 

BEM & 100 (boundary) & 0 & 100 (boundary) & 52 \\[0.5em]

\begin{tabular}{@{}l@{}}3D Simulation\\(COMSOL)\end{tabular} &
\begin{tabular}{@{}c@{}}4208 (boundary)\\16688 (volume)\end{tabular} &
\begin{tabular}{@{}c@{}}6232 (boundary)\\317248 (volume)\end{tabular} &
\begin{tabular}{@{}c@{}}10440 (boundary)\\333936 (volume)\end{tabular} &
1479 \\

\hline\hline
\end{tabular*}
\end{table*}

To compare the simulation methods, we plot the local electric field distribution of the near-field surrounding the apex of the tip. 
The rapid decay of the near field with distance from the tip apex is the physically most relevant parameter for understanding the interaction between tip and sample surface in actual THz sSNOM experiments.
Figures 4(a) and 4(b) display the calculated induced radial $|E_{mag}|$ decay and axial $|E_z|$ decay respectively. The cross-section lines along which the field is plotted for the two simulation types are the family of radial red lines at several constant vertical separations from the tip in Figure~4(a) and the single vertical red line in Figure~4(b) from the apex of the tip downwards.
We clearly see the transition from power-law decay at larger separations to saturation at the vertical-separation scale in the horizontal cross sections, and at the tip-curvature-radius scale in the vertical cross section.
The excellent agreement between the BEM and FEM simulations shows that the BEM method produces valid results. 

Focusing on the radial decay, there are three main decay regions. 
The first is a constant plateau region close to the symmetry axis. 
The radial distance $(\rho)$ is far smaller than the vertical displacement $(d)$ from the surface charges. 
This means that small radial distance shifts produce almost negligible changes in the source-to-observation distance from the induced surface charges, which leads the field strength to be effectively constant. 
The plateau region decreases in magnitude as the vertical displacement $(d)$ increases, as the corresponding radial cut line is farther from the apex. 
The near-field decay begins occurring when $\rho$ becomes comparable to the vertical distance to the apex $(d)$. At this point, the observation-to-source distance is no longer negligible and significantly increases with larger radial distances. 
This is the region that will be highly dependent on the tip geometry as the observation points begin sampling source contributions from broader parts of the tip. 
Radius curvature, cone angle, observation height, and tip height are expected to change the decay shape and strength of field decay. 
Eventually, in the final decay region for distances large compared to the whole tip model, the conductor behaves similarly to an electric dipole, with the decay following $(1/r^3)$. 
The region is shown by the inset of Figure~4(b) as both the radial and axial decay exhibit this behavior.
The region begins to occur when the observation distance is much larger than the spatial extent of the induced charge distribution or tip height.

\section{Conclusion}
In conclusion, the formulated BEM simulation has been shown to produce reliable decay curves while being computationally cheaper and around 28 times faster than 3D FEM simulations. 
This method combines ideas from the Green's function, neutral conductor properties in electrostatics, and power-grading panel spacing to accessibly and efficiently investigate the geometric dependencies of AFM near-field decay. 
Implementing non-axisymmetric tip shapes and electrostatic tip-sample interaction is the immediate next step to prepare the model for realistic probe geometries. 
Ultimately, this method can be used for screening candidate tip geometries that combine different shapes and parameters to meet specific experimental needs.

\begin{acknowledgments}
Work at the Ames National Laboratory was supported by the U.S. Department of Energy Office of Science, Science Undergraduate Laboratory Internships (SULI) program and Basic Energy Science, under its contract with Iowa State University, Contract No. DE-AC02-07CH11358. 
Noah Reich is grateful to the DOE for the assistantship and opportunity to participate in the SULI program.
We thank S.J. Haeuser and R.K. Chan for providing the SEM pictures of the AFM cantilever tip.
\end{acknowledgments}

\bibliography{main}

\end{document}